\documentclass{ws-procs10x7}

\usepackage{amsmath,graphicx}

\begin{document}

\title{Design and Construction of a TPC\\
       using GEM Foils for Gas Amplification}

\author{Peter Wienemann}

\address{Deutsches Elektronen-Synchrotron\\ Notkestr.~85,
         22607 Hamburg, Germany\\E-mail: peter.wienemann@desy.de}

\twocolumn[\maketitle\abstract{
The challenging physics program at the International Linear Collider (ILC)
poses stringent requirements on the performance of its tracking system. A
large volume time projection chamber (TPC) is considered a good candidate for
such a tracker. Whereas conventional TPCs used a wired based gas
amplification system, a future TPC is likely to make use of micro pattern gas
detectors as e.~g.~gas electron multipliers (GEMs) for gas amplification.
This talk gives an overview over recent achievements from the R\&D activities
to build a TPC with a GEM based gas amplification system. This includes charge
transfer studies through multiple GEM structures, field cage design and
spatial resolution measurements in high magnetic fields.}]

\section{Introduction}

The particle physics community recently agreed on building an
$\text{e}^+ \text{e}^-$ linear collider with superconducting
accelerating structures in a joint global effort. The ambitious
physics program at this International Linear Collider (ILC) poses
stringent requirements on the precision of its tracker as part of a
precise overall detector. The measurement of the Higgs properties for
example requires excellent momentum resolution for mass reconstruction
and good particle identification for branching ratio measurements. A
large volume time projection chamber (TPC) as for example proposed for
the TESLA detector\cite{ref:TeslaTDR} is considered a promising
candidate as central tracking device for a detector at ILC. Contrary
to conventional TPCs with a multiwire proportional chamber (MWPC)
technique for gas amplification, future TPCs are likely to make use of
micro pattern gas detectors (MPGDs). The best known representatives of
such MPGDs are gas electron multipliers (GEMs)\cite{ref:GEM} and
micromegas\cite{ref:micromegas}.  MPGDs have amplification structures
of order 100 $\mu$m giving rise to only tiny $\vec{E} \times \vec{B}$
effects, provide a fast and narrow electron signal and have intrinsic
ion feedback suppression -- all features making them good candidates
as gas amplification system of a TPC.

In the following, some of the R\&D activities are described which are carried
out to show that TPCs equipped with GEMs meet the challenging performance
requirements and to prove that they can be operated reliably. The results
presented here were obtained from of a joint R\&D program of various
institutes from around the world.  This linear collider TPC group includes
groups from Aachen, Berkeley, Carleton, Cracow, DESY, Hamburg, Karlsruhe, MIT,
Montreal, MPI Munich, NIKHEF, Novosibirsk, Orsay, St.~Petersburg, Rostock,
Saclay and Victoria.

\section{Charge Transfer through Triple GEM Structures}

\begin{figure}[htb]
\begin{center}
    \includegraphics[width=0.99\columnwidth]{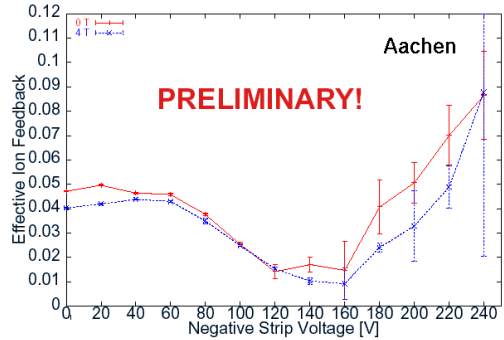}
\end{center}
\caption{The ion backdrift versus the negative voltage applied to the MHSP
  strips.}
\label{fig:mhspionfeedback}
\end{figure}

A crucial item is to understand and to optimize the charge transfer
through GEM structures. The goal is to choose voltage settings which
allow maximal electron transparency to ensure good spatial resolution
and $dE/dx$ accuracy, and in addition minimal ion transparency to keep
drift field distortions due to backdrifting ions small. Charge
transfer parameters like electron/ion collection/extraction efficiency
and gain were measured for various voltage settings from the currents
measured on the various electrodes of a small test chamber being
irradiated by an $^{55}$Fe source. In order to take effects of a
magnetic field into account the measurements were performed in a
superconducting magnet providing fields up to 5 T. The
parametrizations obtained from these measurements allow the
minimization of the ion backdrift (ratio of the cathode and the anode
current) in a convenient way by scanning the available parameter space
using a computer program.
\begin{figure}[htb]
\begin{center}
    \includegraphics[width=0.99\columnwidth]{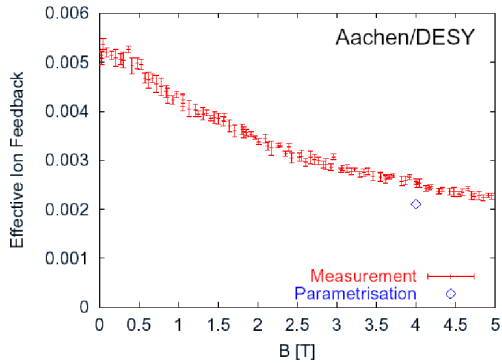}
\end{center}
\caption{The measured ion backdrift versus magnetic field. In addition
  the predicted ion backdrift value from the parametrization is indicated.}
\label{fig:ionbackdrift}
\end{figure}
Figure~\ref{fig:ionbackdrift} shows the measured ion backdrift versus the
magnetic field. The ion backdrift decreases by a factor of two from 0
to 4 T. A value of 0.25 \% is measured and corresponds well with the
prediction from the parametrization. Optimally
\begin{displaymath}
    \mbox{ion backdrift} \times \mbox{gain} < 1,
\end{displaymath}
i.~e.~the number of backdrifting ions from the amplification system is less
than the number of unavoidable primary ions produced in the drift region. For
the achieved ion backdrift this corresponds to running with a gain of less than
400 which is probably not feasible even with new readout electronics
developments. Therefore additional techniques have been attempted to
further suppress ion backdrift. First tests were performed with the first GEM
replaced by a micro hole strip plate (MHSP)\cite{ref:mhsp}. It was
demonstrated in a first proof of principle (see
Fig.~\ref{fig:mhspionfeedback}) that by applying a negative voltage of 150 V
to the MHSP strips the ion backdrift can be reduced by a factor of four. In
order to exploit the full potential of these devices, an optimization similar
to that for GEMs remains to be done.

\section{Field Cage Design}

The field cage has to meet challenging requirements. First of all, it
has to provide a homogeneous electric field in order to avoid track
distortions.  Second, a stable mechanical support structure is needed
to ensure a precise mutual alignment of the various TPC components.
Third, the material budget in terms of radiation lengths has to be
kept small in order to minimize a degradation of the calorimeter
performance. The resistor chain needed to gradually degrade the
potential from the cathode to the anode should dissipate as little
heat as possible into the chamber gas because local temperature
fluctuations change the drift velocity and various other gas
parameters.  Finally the field cage of a large-scale TPC has to stand
cathode voltages of order 50 to 100 kV.

\begin{figure}[htb]
\begin{center}
    \begin{tabular}{c c}
        \begin{minipage}{0.6\columnwidth}
            \includegraphics[width=0.99\columnwidth]{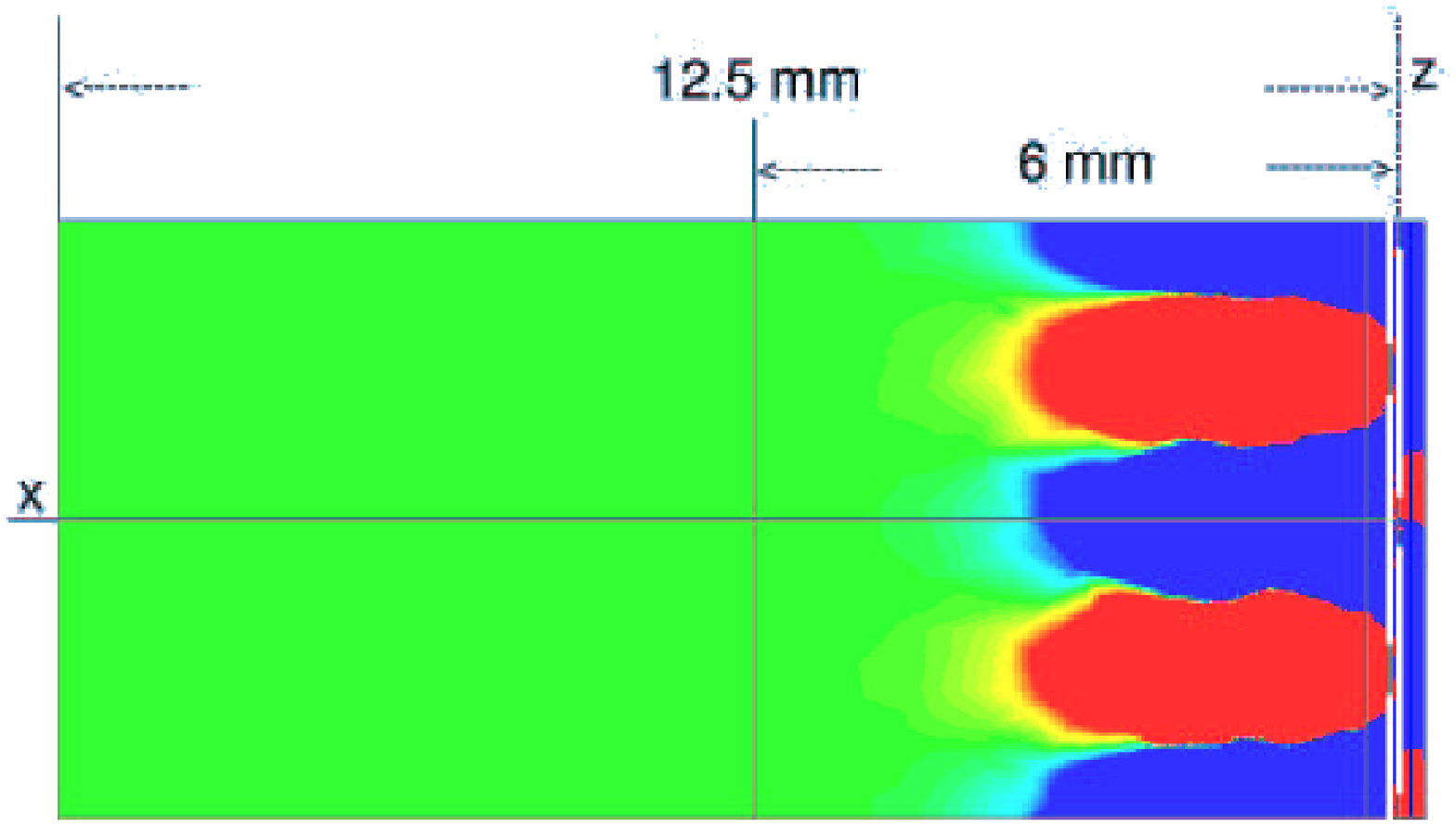}
            \includegraphics[width=0.99\columnwidth]{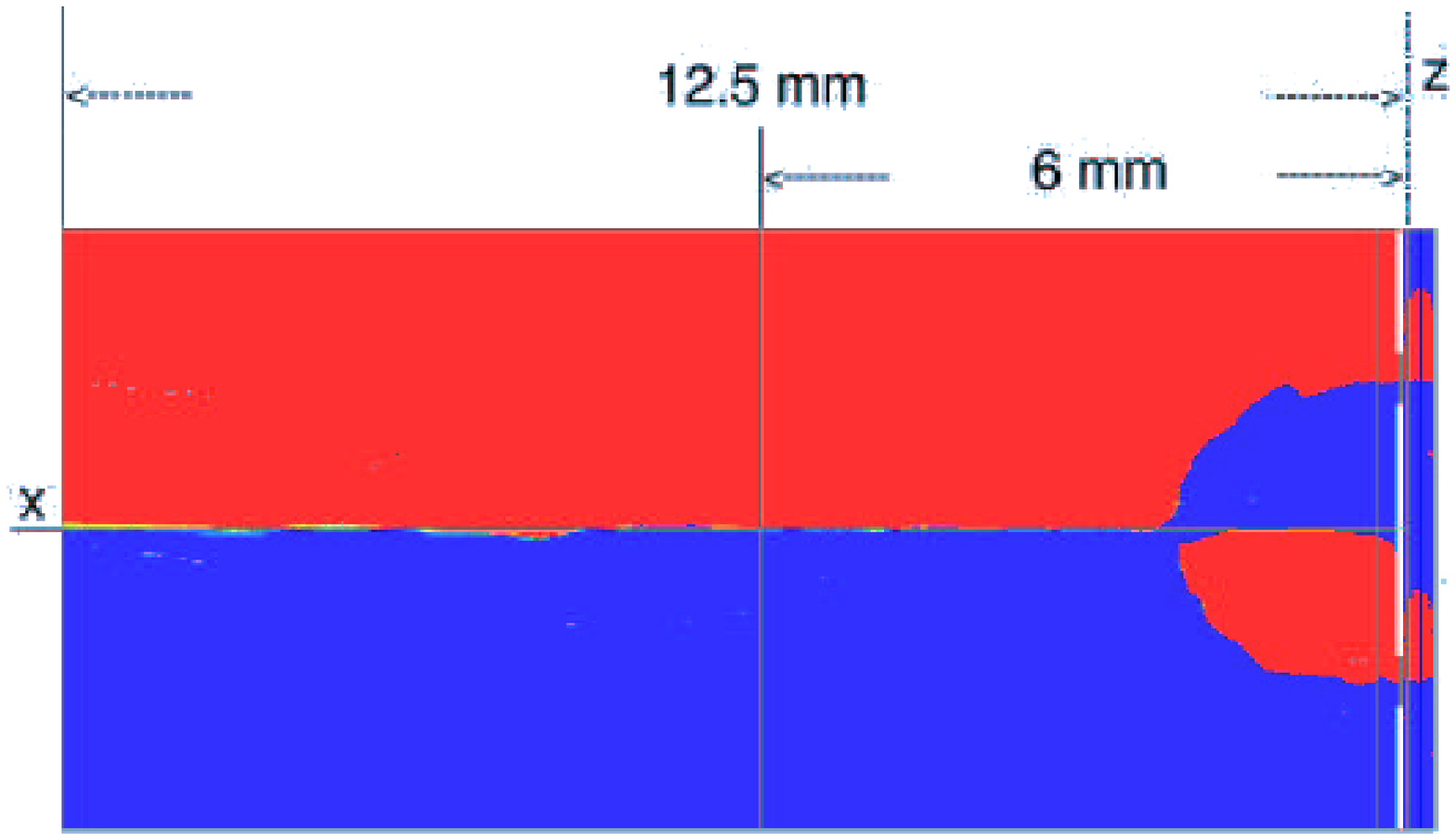}
        \end{minipage}
        &
        \begin{minipage}{0.39\columnwidth}
            \includegraphics[width=0.99\columnwidth]{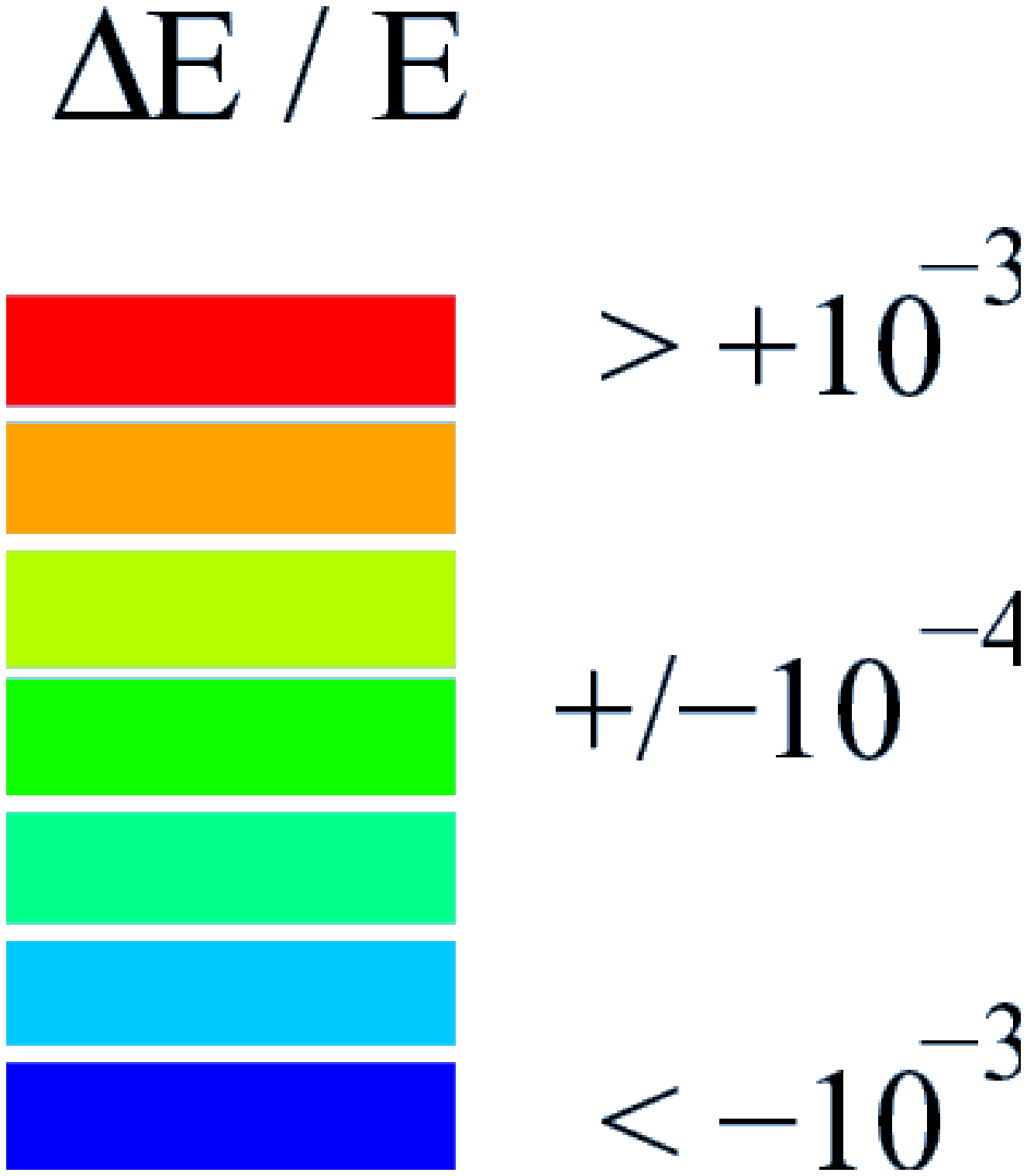}
        \end{minipage}
    \end{tabular}
\end{center}
\caption{The simulated electric field homogeneities for field cages with (top)
  and without (bottom) mirror strips on the outer side of the field cage.}
\label{fig:FCSimulation}
\end{figure}

The electric field homogeneity is mainly determined by the chosen strip
layout. In order to find an optimal setup, simulations have been performed
with the MAXWELL finite element package\cite{ref:maxwell}.
Figure~\ref{fig:FCSimulation} shows the simulated relative $E$ field homogeneity
for field cage designs with and without mirror strips on the outer side of the
field cage. The double-sided strip layout provides inhomogeneities $\Delta E/E
< 10^{-4}$ which is an order of magnitude smaller than what is obtained
without mirror strips on the outside.

Several older TPC prototypes in use by the LC TPC group have only strips on
the inner field cage side. The findings from the simulation study lead to the
construction of a TPC prototype with a double-sided strip field cage. Its
mechanical support structure is composed of honeycomb and glass-fiber
reinforced plastic.  Electrical insulation is provided by four layers of
Kapton. In total the field cage represents only about 1 \% of a radiation
length. It has proven to stand at least 30 kV. To reduce the heat emission of
the resistor chain into the chamber gas, the resistor chain has been placed
outside the gas volume. It is covered by a ceramics plate conducting the
produced heat to the outside and, at the same time, providing good electric
insulation. Following the careful design and test phase, the prototype
performance is currently checked in first measurements with cosmic muons and
a $^{\text{90}}$Sr source.

\section{Transverse Resolution in High Magnetic Fields}

To achieve a high momentum resolution, a good spatial resolution is essential.
The best parameter to compare the performance of different prototypes and to
extrapolate to large-scale devices is the single point resolution. The narrow
MPGD electron signals pose a challenge to accurately reconstruct the track
position with a reasonable number of channels. As opposed to micromegas, GEMs
offer a nice solution. The large diffusion between the individual GEMs of a
multiple GEM structure spreads the charge over a wider area without
sacrificing the track resolution since the defocussing takes place during and
after the gain stage.  The challenge is to find GEM settings and a gas which
provide a good compromise between low diffusion in the drift region and enough
defocussing between the GEMs without severely degrading the two-track
resolution. Several transverse resolution measurements were performed in
magnetic fields up to 5~T both for Ar-CH$_4$ (95-5) and Ar-CH$_4$-CO$_2$
(93-5-2).
\begin{figure}[htb]
\begin{center}
    \includegraphics[width=0.99\columnwidth]{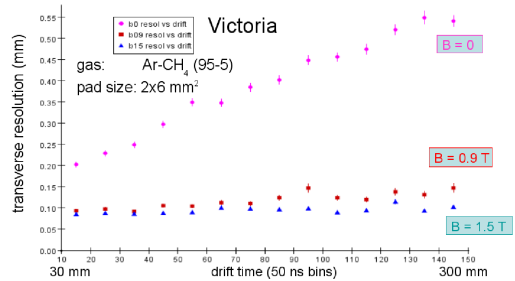}
\end{center}
\caption{The transverse resolution versus drift distance in Ar-CH$_4$
  (95-5).}
\label{fig:resolution}
\end{figure}
Figure~\ref{fig:resolution} shows the results as a function of the
drift distance for Ar-CH$_4$ (95-5) with 2 $\times$ 6 mm$^2$ pads. The spatial
resolution becomes better with increasing magnetic fields since the $B$ field
suppresses diffusion leading to narrower charge distributions arriving at the
gas amplification system.  Already at 1.5~T values below the 100~$\mu$m level
are achieved fulfilling the performance goals mentioned in the TESLA technical
design report\cite{ref:TeslaTDR}.

\section{Conclusion}

The linear collider TPC R\&D activities of the last few years have led to
valuable new insights into the properties and the potential of GEMs as gas
amplification devices in TPCs. Good understanding of the charge transfer
processes in multiple GEM structures has been gained. First important
experiences have been made with building field cages resulting in an
increasing field cage quality in the course of time. Furthermore measurements
performed with such small prototypes revealed that single point resolutions of
the order of 100 $\mu$m are feasible for drift distance below 30~cm with 2
$\times$ 6 mm$^2$ pads. In summary, promising results have been achieved with
small prototypes. Further studies are needed to show that they hold also for
large-scale devices.

\section*{Acknowledgments}
The author would like to thank the members of the linear collider TPC group
for providing their latest results.

\end{document}